\documentclass{article}
\usepackage{spconf,amsmath,graphicx,setspace}
\usepackage{amssymb}
\usepackage{algorithm}
\usepackage{algorithmic}
\usepackage{textcomp}
\usepackage{subfigure}
\usepackage{tipa}

\title{Soft-Label Anonymous Gastric X-ray Image Distillation}
\name{Guang Li $^{\dagger}$ \qquad Ren Togo $^{\dagger\dagger}$ \qquad Takahiro Ogawa $^{\dagger\dagger\dagger}$ \qquad Miki Haseyama$^{\dagger\dagger\dagger}$ \thanks{In our work, the medical data were provided by The University of Tokyo Hospital in Japan. We express our thanks to Nobutake Yamamichi of the Graduate School of Medicine, The University of Tokyo, and Katsuhiro Mabe of the Junpukai Health Maintenance Center. This study was supported in part by the Photo-Excitonix Project of Hokkaido University, and in part by the JSPS KAKENHI under Grant JP17H01744 and Grant JP20K19857.}}
\address{$^{\dagger}$ Graduate School of Information Science and Technology,
    Hokkaido University, Japan \\
    $^{\dagger\dagger}$ Education and Research Center for Mathematical and Data Science,
 	Hokkaido University, Japan \\
    $^{\dagger\dagger\dagger}$ Faculty of Information Science and Technology, 
    Hokkaido University, Japan \\
 	E-mail: \{guang, togo, ogawa\}@lmd.ist.hokudai.ac.jp, miki@ist.hokudai.ac.jp}
\begin{document}
\ninept
\maketitle
%
\begin{abstract}
This paper presents a soft-label anonymous gastric X-ray image distillation method based on a gradient descent approach. 
The sharing of medical data is demanded to construct high-accuracy computer-aided diagnosis (CAD) systems.
However, the large size of the medical dataset and privacy protection are remaining problems in medical data sharing, which hindered the research of CAD systems.
The idea of our distillation method is to extract the valid information of the medical dataset and generate a tiny distilled dataset that has a different data distribution. 
Different from model distillation, our method aims to find the optimal distilled images, distilled labels and the optimized learning rate.
Experimental results show that the proposed method can not only effectively compress the medical dataset but also anonymize medical images to protect the patient’s private information.  
The proposed approach can improve the efficiency and security of medical data sharing.
\end{abstract}
\section{Introduction}
\label{sec:intro}
%
Deep learning, in particular, deep convolutional neural networks (DCNNs), have been popular with many areas in computer vision~\cite{lecun2015deep}.
Especially, in the field of medical image analysis, DCNN-based CAD systems are often used as an auxiliary diagnosis of diseases~\cite{litjens2017survey, faust2018deep}.
The sharing of medical data is a primary method for building high-accuracy CAD systems~\cite{weitzman2010sharing}.
However, there are still two main problems in medical data sharing. 
Firstly, with the increase in the number of medical equipment, the amount of medical image data grows exponentially, resulting in large size of the medical dataset~\cite{goldbeck1996complementary}. 
Therefore, the sharing of medical image data is inefficient.
It is necessary to extract valid data to reduce the size of the medical dataset.
Secondly, because the medical images contain the patient's private information, there is still an enormous controversy over the use of these medical data~\cite{narendra2016medical}.
It is difficult to share the medical image data of many diseases, which hindered the research and development of CAD systems.  
\par
To solve the problem of reducing the size of the dataset, many researchers have proposed unique solutions.
For example, many instance selection and dataset pruning methods aim to select a subset of the entire training dataset and achieve performance comparable to the original full dataset, which can reduce the size of the dataset~\cite{olvera2010review, bachem2017practical}.
Similarly, Campbell $et\,al$. proposed a Bayesian core-set construction method that can find valid data in the original dataset via the greedy iterative geodesic ascent~\cite{campbell2018bayesian}.   
Also, Sener $et\,al$. proposed a core-set construction approach based on active learning~\cite{sener2017active}. 
Active learning reduces the amount of data that need to be labeled by only labeling data that are difficult to classify~\cite{cohn1996active, tong2001support}.
Although these previous approaches have made some progresses in solving the problem of reducing the size of standard image dataset, due to the high complexity of medical images and the need for professional knowledge of labeling, they still cannot meet the requirements for medical dataset.
\par
Privacy protection problem has always been the main obstacle in medical data sharing~\cite{mcgraw2009privacy, malin2010technical}. 
Researchers have also tackled the problem of hiding personal information. 
For example, some methods for removal of identifiers have been proposed to protect the privacy of patients~\cite{berman2002confidentiality}.
These methods contain stripping identifier and one-way hashing algorithm.
The stripping identifier approach is an anonymization algorithm which can remove all patient identifiers in a record. 
And the one-way hashing algorithm transforms a patient record string into another string, such that operations on the hash value cannot calculate the original patient record. 
In recent years, with the rise of big data, some ways to use cloud computing platforms to share medical data securely have been proposed.
For example, Yang $et\,al$. proposed a hybrid solution method of privacy-preserving medical data sharing based on cloud computing~\cite{yang2015hybrid}.
Also, Fabian $et\,al$. proposed a method to realize the safe sharing of medical data with semi-trust cloud computing environments~\cite{fabian2015collaborative}.
Since it is challenging to ensure the validity of anonymized medical images, all of these methods do not pay attention to the anonymization of medical image data itself.
\par
In this paper, we propose a novel method that can not only effectively compress the medical dataset but also anonymize medical images to protect the patient's private information~\cite{bengio2000gradient, maclaurin2015gradient}.
Since the researches of gastric X-ray images have both of the above problems, and we have previously proposed methods for anonymous gastritis image generation and automatic detection of gastritis~\cite{togo2018anonymous, kanai2019gastritis}, we also focus on gastric X-ray images in this research.
Considering the gastric X-ray images with high resolutions can lead to expensive computing cost, we divide them into patches that have three categories. 
To maximize the compression of the dataset and distillation of the valid information, we distill each class into one image for training.
Then we use the distilled images to estimate the labels of full gastric X-ray images.
Experimental results show that our distillation method achieved competitive classification accuracy using a tiny distilled dataset.
Furthermore, the sharing of medical image data can become more efficient and safer because the distilled images do not contain private information.
\par
Our contributions are summarized as follows: 
\begin{itemize}
    \item We propose a novel method for anonymous medical image distillation, which can improve the efficiency and security of medical image data sharing.
    \item We realize high classification performance by distilling each class into one image for training.
\end{itemize}

\section{Anonymous Gastric X-ray image distillation}
This section shows the details of the soft-label anonymous gastric X-ray image distillation method.
In subsection 2.1, we preprocess the training dataset by dividing the gastric images into patches.
In subsection 2.2, we demonstrate the entire flow of the anonymous gastric image distillation algorithm.
In subsection 2.3, we explain how to estimate the labels of full gastric X-ray images.
\subsection{Patch-based Gastric X-ray Image Labeling}
\begin{figure}[!ht]
        \centering
        \subfigure[]{
        \centering
        \includegraphics[width=4.0cm]{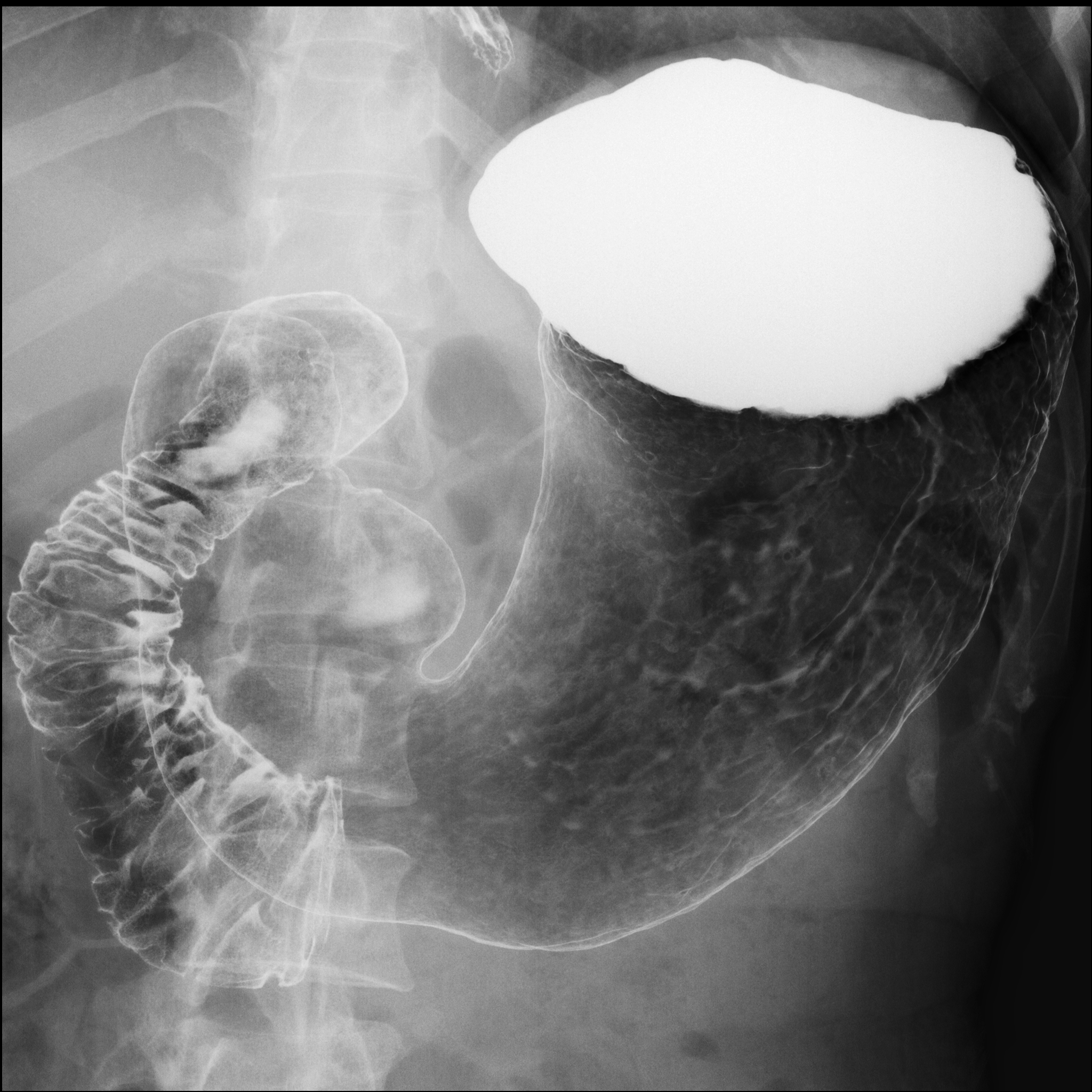}
        }
        \subfigure[]{
        \centering
        \includegraphics[width=4.0cm]{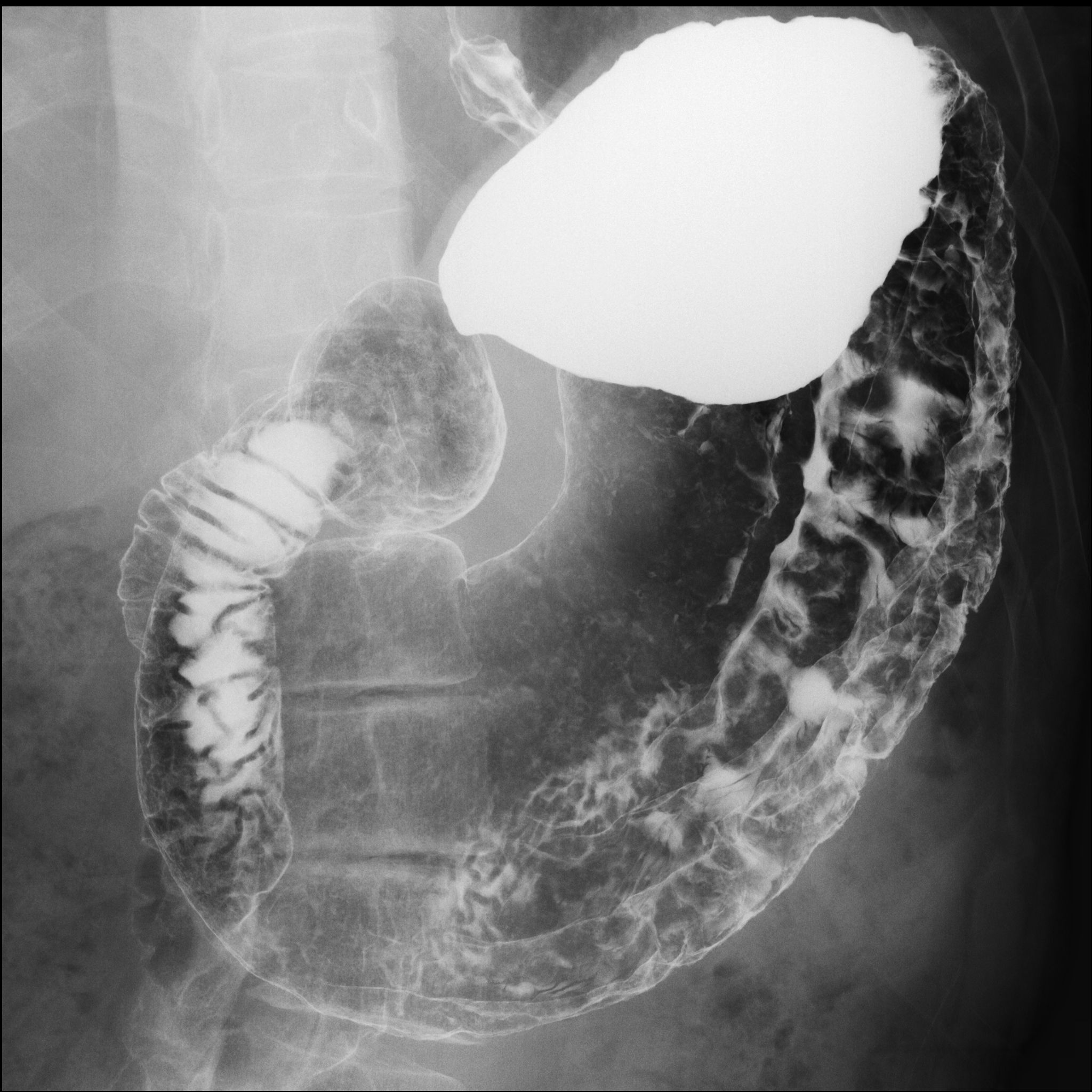}
        }
        \caption{Examples of full gastric X-ray images: (a) a sample of non-gastritis image, (b) a sample of gastritis image.}
        \label{fig1}
\end{figure}

\begin{figure}[!ht]
        \centering
        \subfigure[]{
        \centering
        \includegraphics[width=1.6cm]{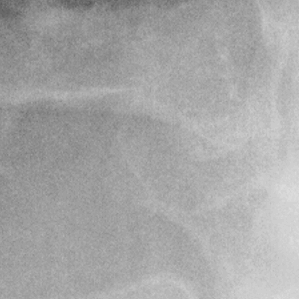}
        \includegraphics[width=1.6cm]{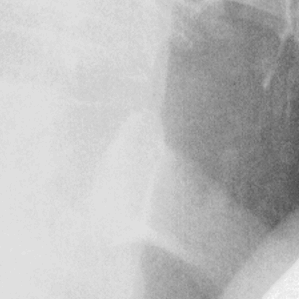}
        \includegraphics[width=1.6cm]{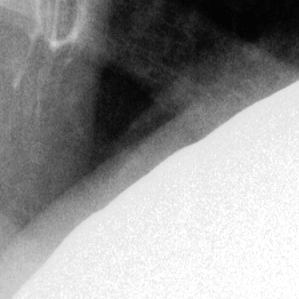}
        \includegraphics[width=1.6cm]{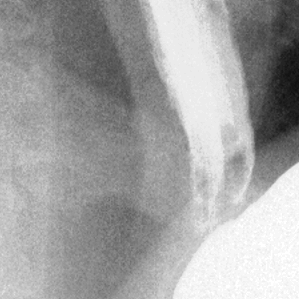}
        \includegraphics[width=1.6cm]{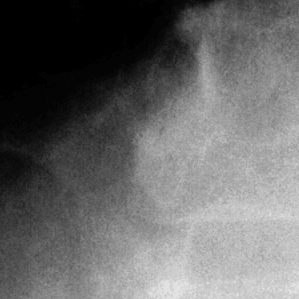}
        }
        \subfigure[]{
        \centering
        \includegraphics[width=1.6cm]{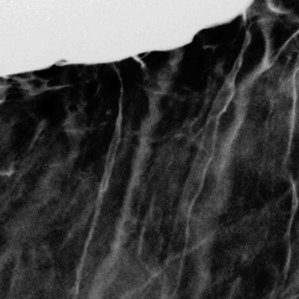}
        \includegraphics[width=1.6cm]{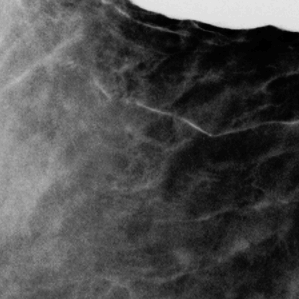}
        \includegraphics[width=1.6cm]{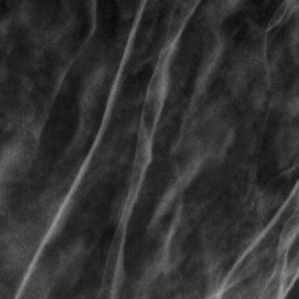}
        \includegraphics[width=1.6cm]{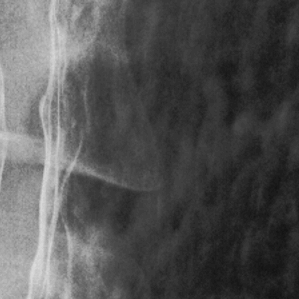}
        \includegraphics[width=1.6cm]{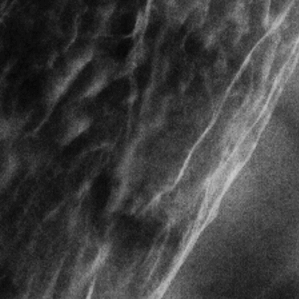}
        }
        \subfigure[]{
        \centering
        \includegraphics[width=1.6cm]{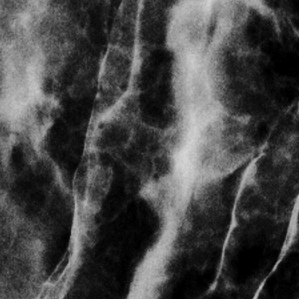}
        \includegraphics[width=1.6cm]{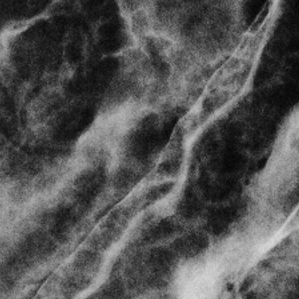}
        \includegraphics[width=1.6cm]{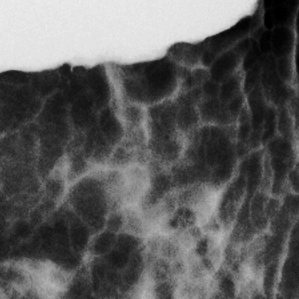}
        \includegraphics[width=1.6cm]{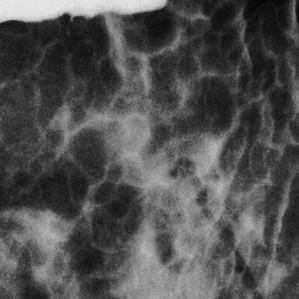}
        \includegraphics[width=1.6cm]{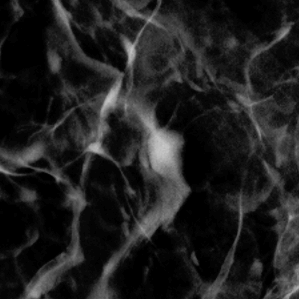}
        }
        \caption{Examples of patch-based gastric X-ray images: (a) irrelevant patches in $\mathcal{I}$, (b) non-gastritis patches in $\mathcal{N}$, (c) gastritis patches in $\mathcal{P}$.}
        \label{fig2}
\end{figure}

This subsection shows how we preprocess the training dataset.
The gastric X-ray images in our dataset have high resolutions, $e.g.$, 2,048 $\times$ 2,048 pixels.
In practical applications, the high-resolution images can lead to expensive computing cost.
Therefore, we divide each gastric X-ray image into $H \times W$ patches ($H$ and $W$ respectively denote the number of patches in the vertical direction and the horizontal direction) and manually label these patch-based gastric X-ray images into the following three categories: 
\begin{itemize}
    \item $\mathcal{I}$: patches outside of the stomach (irrelevant), 
    \item $\mathcal{N}$: patches extracted from non-gastritis X-ray images (negative) inside of the stomach,
    \item $\mathcal{P}$: patches extracted from gastritis X-ray images (positive) inside of the stomach. 
\end{itemize}
Figures \ref{fig1} and \ref{fig2} show the full gastric X-ray images and patch-based gastric X-ray images, respectively.
\begin{algorithm*}[t]
    \caption{Anonymous Gastric Image Distillation}    
    \label{alg1}
    \begin{algorithmic}[1]
    \REQUIRE 
    $\theta_{0}$: the random initial weights;
    $M$: the number of distilled images;
    $E$: distill epochs;
    $I$: distill steps;
    \\
    \ \ \ \ \ \ $\alpha$: learning rate;
    $K$: batch size;
    $T$: training steps;
    $\tilde{\mathbf{y}}_{0}$: initial value for $\tilde{\mathbf{y}}$;
    $\tilde{\alpha}_{0}$: initial value for $\tilde{\alpha}$
    \ENSURE
    $\tilde{\mathbf{x}}$: distilled images;
    $\tilde{\mathbf{y}}$: distilled labels;
    $\tilde{\alpha}$: optimized learning rate
    \\
    \STATE
    Initialize 
    $\tilde{\mathbf{x}}$ = $\left \{ \tilde{x}_{m} \right \}_{m=1}^{M}$ randomly,
    $\tilde{\mathbf{y}}$ = $\left \{ \tilde{y}_{m} \right \}_{m=1}^{M}\leftarrow\tilde{\mathbf{y}}_{0}$,
    $\tilde{\alpha}\leftarrow\tilde{\alpha}_{0}$
    \FOR{each training step $t = 1$ to $T$}
    \STATE
    Get a minibatch of training data $\left ( \mathbf{x}_{t}, \mathbf{y}_{t} \right )$ = $\left \{ x_{t,k}, y_{t,k} \right \}_{k=1}^{K}$
    \STATE
    Get the random initial weights $\theta_{0}$
    \FOR{each distilling epoch $e = 1$ to $E$} 
    \FOR{each distilling step $i = 0$ to $I - 1$} 
    \STATE
    Compute updated weights with gradient descent:
    $\theta_{i+1} \leftarrow \theta_{i} - \tilde{\alpha} \, \nabla_{\theta_{i}} \ell \left ( \tilde{\mathbf{x}}, \tilde{\mathbf{y}}, \theta_{i} \right)$ 
    \STATE
    Evaluate the objective function on the minibatch of training data:
    $\mathcal{L} = \ell \left ( \mathbf{x}_{t}, \mathbf{y}_{t}, \theta_{i+1}  \right )$ 
    \STATE
    Update distilled data:
    $\tilde{\mathbf{x}} \leftarrow \tilde{\mathbf{x}} - \alpha \nabla_{\tilde{\mathbf{x}}} \mathcal{L}$,
    $\tilde{\mathbf{y}} \leftarrow \tilde{\mathbf{y}} - \alpha \nabla_{\tilde{\mathbf{y}}} \mathcal{L}$,
    and $\tilde{\mathbf{\alpha}} \leftarrow \tilde{\mathbf{\alpha}} - \alpha \nabla_{\tilde{\mathbf{\alpha}}} \mathcal{L}$ 
    \ENDFOR
    \ENDFOR
    \ENDFOR
    \end{algorithmic}
\end{algorithm*}
\subsection{Anonymous Gastric Image Distillation}
In this subsection, we explain our anonymous gastric image distillation algorithm.
When we have the patch-based gastric training dataset $\left ( \mathbf{x, y} \right )$ = $\left \{ x_{g}, y_{g} \right \}_{g=1}^{G}$, where $G$ denotes the number of training images, $x_{g}$ and $y_{g}$ denote the image and the corresponding label, respectively.
We parameterize a DCNN model as $\theta$, and let the twice-differentiable loss function $\ell \left ( \mathbf{x, y}, \theta \right )$ denotes the loss of this network on the entire training dataset $\left ( \mathbf{x, y} \right )$.
Considering the gastritis patches and non-gastritis patches may have common features, we let the distilled images $\tilde{\mathbf{x}}$ have soft labels $\tilde{\mathbf{y}}$, which contain $\mathcal{I}$, $\mathcal{N}$ and $\mathcal{P}$~\cite{sucholutsky2019soft}.
In our gastric image distillation method, we distill the valid information of the entire training dataset to a tiny distilled dataset $\left ( \tilde{\mathbf{x}}, \tilde{\mathbf{y}} \right)$ = $\left \{ \tilde{x}_{m}, \tilde{y}_{m} \right \}_{m=1}^{M}$ with $M \ll G$.
For each distilling step $i$, the weights are updated as follows:
\begin{equation}
\label{equ1}
\theta_{i+1} \leftarrow \theta_{i} - \tilde{\alpha} \nabla_{\theta_{i}} \ell 
\left ( \tilde{\mathbf{x}}, \tilde{\mathbf{y}}, \theta_{i} \right),
\end{equation}
where $\left ( \tilde{\mathbf{x}}, \tilde{\mathbf{y}} \right)$ denotes the distilled dataset, and $\tilde{\alpha}$ denotes the optimized learning rate.
%
%
%
%
\par
With the derived new weights $\theta_{i+1}$ of the DCNN model, we evaluate it on the entire training dataset $\left ( \mathbf{x, y} \right )$.
Our goal is to find the optimal distilled images $\tilde{\mathbf{x}}^{*}$, distilled labels $\tilde{\mathbf{y}}^{*}$ and the optimized learning rate $\tilde{\mathbf{\alpha}}^{*}$.
The objective function is defined as:
\begin{equation}
\label{equ2}
\begin{split}
\tilde{\mathbf{x}}^{*}, \tilde{\mathbf{y}}^{*}, \tilde{\mathbf{\alpha}}^{*} & = 
\mathrm{arg \, min} \,\mathcal{L}\left ( \tilde{\mathbf{x}}, \tilde{\mathbf{y}}, \tilde{\alpha}; \theta_{i} \right ), \\ & = 
\mathrm{arg \, min} \, \ell \left ( \mathbf{x, y,} \, \theta_{i+1} \right ), \\ & =
\mathrm{arg \, min} \, \ell \left ( \mathbf{x, y,} \, \theta_{i} - \tilde{\alpha} \nabla_{\theta_{i}} \ell \left ( \tilde{\mathbf{x}}, \tilde{\mathbf{y}}, \theta_{i} \right) \right ), 
\end{split}
\end{equation}
where $\ell \left ( \tilde{\mathbf{x}}, \tilde{\mathbf{y}}, \theta_{i} \right)$ is twice-differentiable, and $\mathcal{L}\left ( \tilde{\mathbf{x}}, \tilde{\mathbf{y}}, \tilde{\alpha}; \theta_{i} \right )$ is differentiable.
\par
To obtain the optimal distilled images, distilled labels and the optimized learning rate, we update the distilled images $\tilde{\mathbf{x}}$, distilled labels $\tilde{\mathbf{y}}$ and the optimized learning rate $\tilde{\mathbf{\alpha}}$ at each distilling step with gradient descent as follows:
\begin{equation}
\label{equ3}
\begin{split}
&
\tilde{\mathbf{x}} \leftarrow \tilde{\mathbf{x}} - \alpha \nabla_{\tilde{\mathbf{x}}} \mathcal{L}, \\ &
\tilde{\mathbf{y}} \leftarrow \tilde{\mathbf{y}} - \alpha \nabla_{\tilde{\mathbf{y}}} \mathcal{L}, \\ &
\tilde{\mathbf{\alpha}} \leftarrow \tilde{\mathbf{\alpha}} - \alpha \nabla_{\tilde{\mathbf{\alpha}}} \mathcal{L},
\end{split}
\end{equation}
where $\nabla_{\tilde{\mathbf{x}}} \mathcal{L}$, $\nabla_{\tilde{\mathbf{y}}} \mathcal{L}$ and $\nabla_{\tilde{\mathbf{\alpha}}} \mathcal{L}$ respectively denote the gradient of $\mathcal{L}$ based on $\tilde{\mathbf{x}}$, $\tilde{\mathbf{y}}$ and $\tilde{\mathbf{\alpha}}$, $\alpha$ denotes the learning rate.
\par
Then we demonstrate the details of the anonymous gastric image distillation algorithm.  
Firstly, we show our input and output settings.
Let $\theta_{0}$ denotes the random initial weights of the DCNN model, $M$, $E$ and $I$ respectively denote the number of distilled images, distill epochs and distill steps.
Also, let $\alpha$, $K$ and $T$ are respectively denote learning rate, batch size and training steps. 
In addition, $\tilde{\mathbf{y}}_{0}$ is the initial value of $\tilde{\mathbf{y}}$, and $\tilde{\alpha}_{0}$ is the initial value of $\tilde{\alpha}$.
And we obtain the distilled images $\tilde{\mathbf{x}}$, distilled labels $\tilde{\mathbf{y}}$ and the optimized learning rate $\tilde{\alpha}$ on the output.
\par
Next, we explain the training process of the algorithm.
First, we randomly initialize the distilled images $\tilde{\mathbf{x}}$.
Then we initialize the distilled labels $\tilde{\mathbf{y}}$ and the optimized learning rate $\tilde{\alpha}$  with $\tilde{\mathbf{y}}_{0}$ and $\tilde{\alpha}_{0}$, respectively. 
At each training step $t$, we get a minibatch of training data $\left ( \mathbf{x}_{t}, \mathbf{y}_{t} \right )$ whose size is $K$.
Then we get the random initial weights $\theta_{0}$ and begin the distilling process.
We repeat the distilling process for $E$ times.
At each distilling step $i$, we compute the updated weights with Eq.~(\ref{equ1}). 
Then we evaluate the objective function on the minibatch of training data with Eq.~(\ref{equ2}).
Finally, we update the distilled data $\tilde{\mathbf{x}}$, $\tilde{\mathbf{y}}$, and $\tilde{\alpha}$ with Eq.~(\ref{equ3}) based on gradient descent~\cite{bergstra2011algorithms, finn2017model}. 
Algorithm~\ref{alg1} shows the entire flow of our proposed approach.
\subsection{Full Gastric X-ray Image Classification}
In this subsection, we explain how to estimate the label of a full gastric X-ray image based on patches.
First, when we have a test gastric image, we divide it into $H \times W$ patches.
Then we put the divided patches into the trained DCNN model, and we can obtain the predicted labels of these patches.
Next, we respectively calculate the number of patches whose predicted labels are $\mathcal{N}$ and $\mathcal{P}$.
The $\mathcal{I}$ patches that extracted from outside of the stomach are not related to the gastritis/non-gastritis prediction, and hence we do not take into account them to the probability calculation.  
Finally, we estimate the label of a full gastric X-ray image as follows:
\begin{equation}
y^\mathrm{test} = 
\left\{\begin{matrix}
 1 & \mathrm{if} \, \frac{\mathrm{Num}(\mathcal{P})}{\mathrm{Num}(\mathcal{P})+\mathrm{Num}(\mathcal{N})} \geq \epsilon \hfill \\
 0 & \mathrm{otherwise} \hfill
\end{matrix}\right.
,
\end{equation}
where $\epsilon$ is a threshold, $\mathrm{Num}(\mathcal{P})$ denotes the number of gastritis patches, and $\mathrm{Num}(\mathcal{N})$ denotes the number of non-gastritis patches.
Note that if $y^\mathrm{test} = 1$, the estimation result of a full gastric X-ray image is gastritis, and the estimation result is non-gastritis otherwise.
\section{Experimental Results}
In this section, we verify the effectiveness of our distillation method with experimental results.
In subsection 3.1, we show the experimental settings of our method.
In subsection 3.2, we evaluate the performance of our method by classifying the full gastric X-ray images.
\subsection{Experimental Settings}
This subsection shows the experimental settings of our research. 
The dataset used in our study contains 815 patients' (240 gastritis and 575 non-gastritis) gastric X-ray images.
Each image has a ground truth (gastritis/non-gastritis), which was determined by patient diagnosis results of endoscopic examination and X-ray inspection. 
All of the gastric X-ray images are gray-scale and high-resolution. 
The training dataset contains 200 patients' (100 gastritis and 100 non-gastritis) images.
Also, the rest of the patients' (140 gastritis and 475 non-gastritis) images are included in the test dataset.
In the data preprocessing stage, we divided the images into patches (299 $\times$ 299 pixels), $i.e.$, $H$ = $W$ = 35, where the sliding interval was set to 50 pixels.
Besides, the patches extracted from the training dataset were labeled as $\mathcal{I}$, $\mathcal{N}$ and $\mathcal{P}$ by a radiological technologist.
Note that if the regions inside of the stomach were less than 1$\%$ in a patch, it was labeled as $\mathcal{I}$.
In addition, if the regions inside of the stomach were more than 85$\%$ in a patch, it was labeled as $\mathcal{N}$ or $\mathcal{P}$.
And we discarded the rest of the patches in the training dataset.
As a result, we obtained $\mathcal{I}$, $\mathcal{N}$ and $\mathcal{P}$ whose number of patches were 48,385, 42,785 and 45,127, respectively.
\par
In the training phase, we loaded the patch-based training dataset ($\mathcal{I}$, $\mathcal{N}$ and $\mathcal{P}$) into a DCNN model.
Since the distilling process involves the calculation of a quadratic gradient, a complicated DCNN model can lead to inefficiency.
Hence, we constructed the ResNet18~\cite{he2016deep} model with PyTorch framework~\cite{paszke2017automatic, paszke2019pytorch}, and the loss $\ell$ was cross entropy loss.
In our experiments, we set the number of distilled images of each category to 1, $i.e.$, $M$ = 3.
Besides, the distill epochs and steps were set to 3, $i.e.$, $E$ = $I$ = 3 (total 9 distill steps).
And we initialized the soft labels with one-hot values of the original labels, which tends to have higher performance compared to the random initialization.
The hard-label distillation method had the same settings as soft-label distillation, except for the labels being fixed~\cite{wang2018dataset}.   
As a result, the training process distilled each class into one image and saved the distilled images, distilled labels and the optimized learning rate.
We performed 400 epochs in both of soft-label distillation and hard-label distillation and saved the results after every ten epochs for testing and evaluation.
It is difficult for a DCNN model to learn from only several images, so we respectively selected 1,000, 2,000 and 3,000 images per category from the patch-based training dataset and trained three ResNet18 models until convergence.
\par
In the test phase, we performed two experiments (Ex. I and Ex. I\hspace{-.1em}I) to show the validity of our proposed method quantitatively.  
Firstly, in Ex. I, we selected the distilled soft-label images that have the best classification performance on the patch-based training dataset and evaluated the performance on the full X-ray images of the test dataset. 
Specifically, we evaluated the full gastric X-ray images classification performance of the ResNet18 models trained on the three random subsets.
Secondly, in Ex. I\hspace{-.1em}I, we used the hard-label distillation as a comparative method.
As in Ex. I, we selected the best distilled hard-label images, which have the highest classification accuracy on the patch-based training dataset.
And we utilized the distilled hard-label images to evaluate the classification performance on the full gastric X-ray images.
Note that we set the threshold $\epsilon$ to 0.4 in both of Exs. I and I\hspace{-.1em}I, which tends to have an excellent classification performance.
We utilized the following sensitivity (Sen), specificity (Spe) and harmonic mean (HM) of Sen and Spe as evaluation indexes:
\\
\begin{equation}
\mathrm{Sen} = \frac{\mathrm{TP}}{\mathrm{TP + FN}},
\end{equation}
\begin{equation}
\mathrm{Spe} = \frac{\mathrm{TN}}{\mathrm{TN + FP}},
\end{equation}
\begin{equation}
\mathrm{HM} = \frac{\mathrm{2 \times Sen \times Spe}}{\mathrm{Sen + Spe}},
\end{equation}
\\
where TP, FN, TN and FP denote the number of true positive, false negative, true negative and false positive, respectively.  
\begin{table}[t]
    \small
    \centering
    \caption{Results of proposed method compared with ResNet18 trained on the random subsets in Ex. I.}
    \label{tab1}
    \begin{tabular}{lccc}
    \\
    \hline
    Method & Sen & Spe & HM \\\hline\hline
    \bfseries{Proposed Method} & 0.886 & \bfseries{0.869} & \bfseries{0.877} \\\hline
    ResNet18 (3000) & 0.814 & 0.832 & 0.823 \\\hline
    ResNet18 (2000) & 0.907 & 0.760 & 0.827 \\\hline
    ResNet18 (1000) & \bfseries{0.914} & 0.669 & 0.773 \\
    \hline
    \end{tabular}
\end{table}
\begin{table}[t]
    \small
    \centering
    \caption{Results of proposed method compared with hard-label distillation in Ex. I\hspace{-.1em}I.}
    \label{tab2}
    \begin{tabular}{lccc}
    \\
    \hline
    Method & Sen & Spe & HM \\\hline\hline
    \bfseries{Proposed Method} & \bfseries{0.886} & 0.869 & \bfseries{0.877} \\\hline 
    Hard-Label Distillation & 0.829 & \bfseries{0.884} & 0.856 \\
    \hline
    \end{tabular}
\end{table}
\begin{figure}[t]
        \centering
        \subfigure[distilled hard-label images at distill step 9]{
        \centering
        \includegraphics[width=8.3cm]{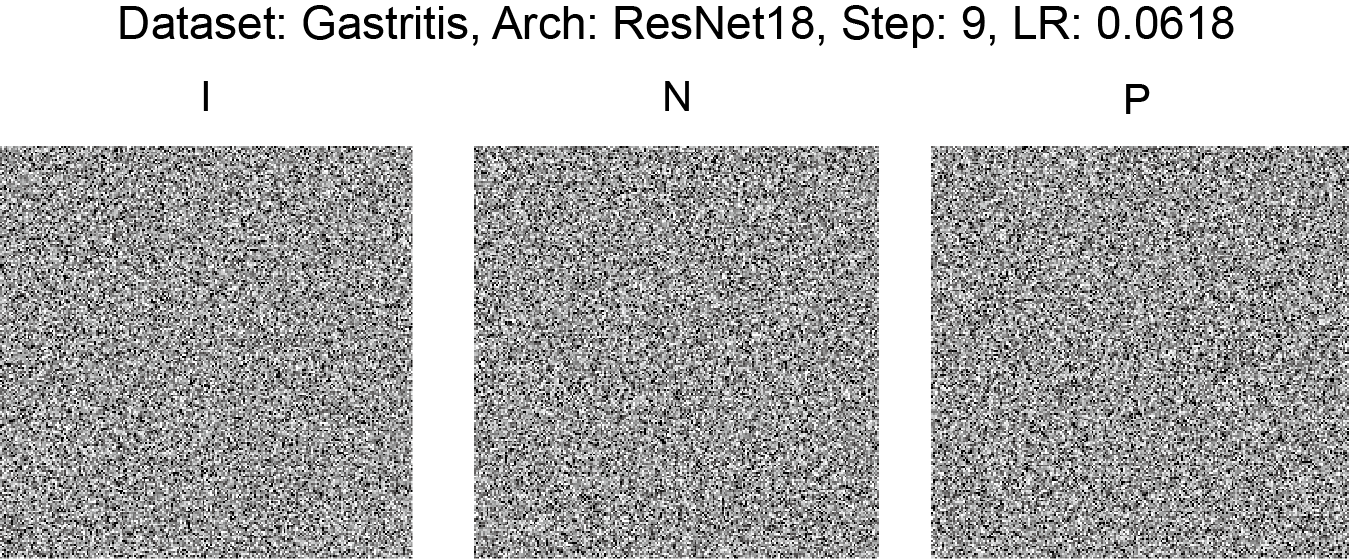}
        }
        \subfigure[distilled soft-label images at distill step 9]{
        \centering
        \includegraphics[width=8.3cm]{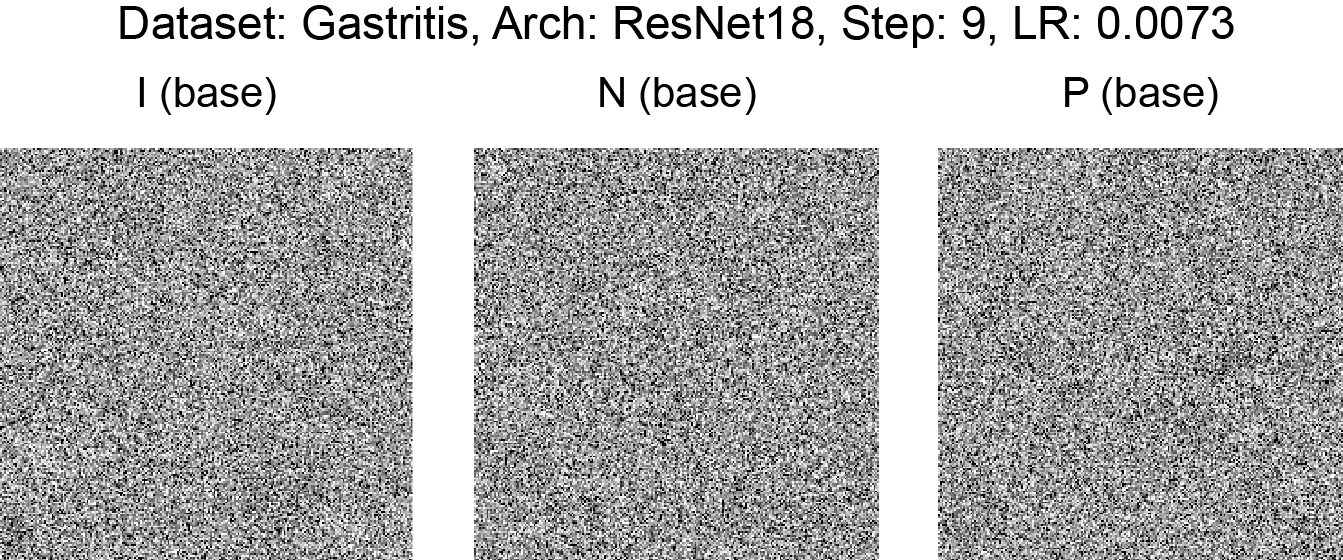}
        }
        \caption{Examples of distilled images. I: the distilled image of label $\mathcal{I}$, N: the distilled image of label $\mathcal{N}$, P: the distilled image of label ${\mathcal{P}}$, LR: the optimized learning rate.}
        \label{fig3}
\end{figure}
\subsection{Results and Discussion}
The experimental results are shown in Tables \ref{tab1} and \ref{tab2}.
Table \ref{tab1} shows the full gastric X-ray images classification performance of our proposed method and ResNet18 trained on the random subsets. 
The ResNet18 model trained with 3,000 images per category (total 9,000 images) has an HM score of 0.823.
On the other hand, our soft-label distillation method that distilled each class into only one image for training has a higher HM score of 0.877.
We can see that the proposed method realizes competitive classification accuracy with a tiny distilled dataset.
Table \ref{tab2} shows the full gastric X-ray images classification performance of our soft-label distillation and the hard-label distillation method.  
The HM scores of both the two approaches exceed 0.85, but our soft-label distillation outperforms the hard-label distillation with a higher score.
Furthermore, we can see that the classification performance of our method becomes more stable because of the balance of Sen score and Spe score.  
It means that the distilled images with soft-label can lead to better distillation results.
Experimental results clearly showed the validity of our purposed soft-label distillation method.
\par
Figure \ref{fig3} shows examples of distilled hard-label images and soft-label images used in our experiments.
With the distillation methods, the information of gastritis/non-gastritis patch images was extracted and merged into only one image.
Note that I (base), N (base) and P (base) respectively denote the image belongs to label $\mathcal{I}$, $\mathcal{N}$ and $\mathcal{P}$ with the highest probability.
From Fig. \ref{fig3}, we can see that the features of gastritis/non-gastritis patches cannot be distinguished, in other words, the gastric images were anonymized completely.
Hence, the distilled images have no private information of patients.
Figure \ref{fig3} clearly showed that our proposed method can effectively compress and anonymize the medical image data.
\section{CONCLUSION}
In this paper, we have proposed a soft-label anonymous gastric image distillation method.
The proposed method realizes high classification performance by distilling each class into one image for training.
Furthermore, the sharing of medical image data with our approach can become more efficient and safer because the distilled images are anonymous.
%

\newpage
\bibliographystyle{IEEEbib}
\bibliography{refs}

\end{document}